\documentclass{iau}
\usepackage{graphicx}

\title[Surface differential rotation of IL\,Hya] 
{Surface differential rotation of IL\,Hya from time-series Doppler images}

\author[ Zsolt K\H{o}v\'ari et al.]   
{Zsolt K\H{o}v\'ari$^1$,
 Levente Kriskovics$^1$,
 Katalin Ol\'ah$^1$
 Kriszti\'an Vida$^1$
 J\'anos Bartus$^1$,
 Klaus G. Strassmeier$^2$,
 \and Michael Weber$^2$}

\affiliation{Konkoly Observatory, \\
$^1$Konkoly Thege \'ut 15-17., H-1121, Budapest, Hungary \\ email: {\tt kovari}, {\tt kriskovics}, {\tt olah}, {\tt vida}, {\tt bartus@konkoly.hu} \\[\affilskip]
$^2$ Leibniz Institute for Astrophysics Potsdam, \\ 
An der Sternwarte 16, 14482 Potsdam, Germany \\ email: {\tt kstrassmeier}, {\tt mweber@aip.de} }

\pubyear{2013}
\volume{302}  
\pagerange{100--101}
\jname{Magnetic Fields Throughout the Stellar Evolution}
\editors{A.C. Editor, B.D. Editor \& C.E. Editor, eds.}
\begin{document}

\maketitle

\begin{abstract}
We present a time-series Doppler imaging study of the K-subgiant component in the RS\,CVn-type binary system IL\,Hya ($P_{\rm orb}=12.905$\,d).
From re-processing the unique long-term spectroscopic dataset of 70 days taken in 1996/97, we perform a thorough cross-correlation analysis
to derive surface differential rotation. As a result we get solar-type differential rotation with a shear value $\alpha$ of 0.05,
in agreement with preliminary suggestions from previous attempts. A possible surface pattern of meridional circulation is also detected.

\keywords{stars: activity,
    stars: imaging,
    stars: individual (IL\,Hya),
    stars: spots,
    stars: late-type}
\end{abstract}

\firstsection 
\section{Time-series Doppler images of IL\,Hya}

\begin{figure}[b]
\begin{center}
 \includegraphics[height=3.0cm]{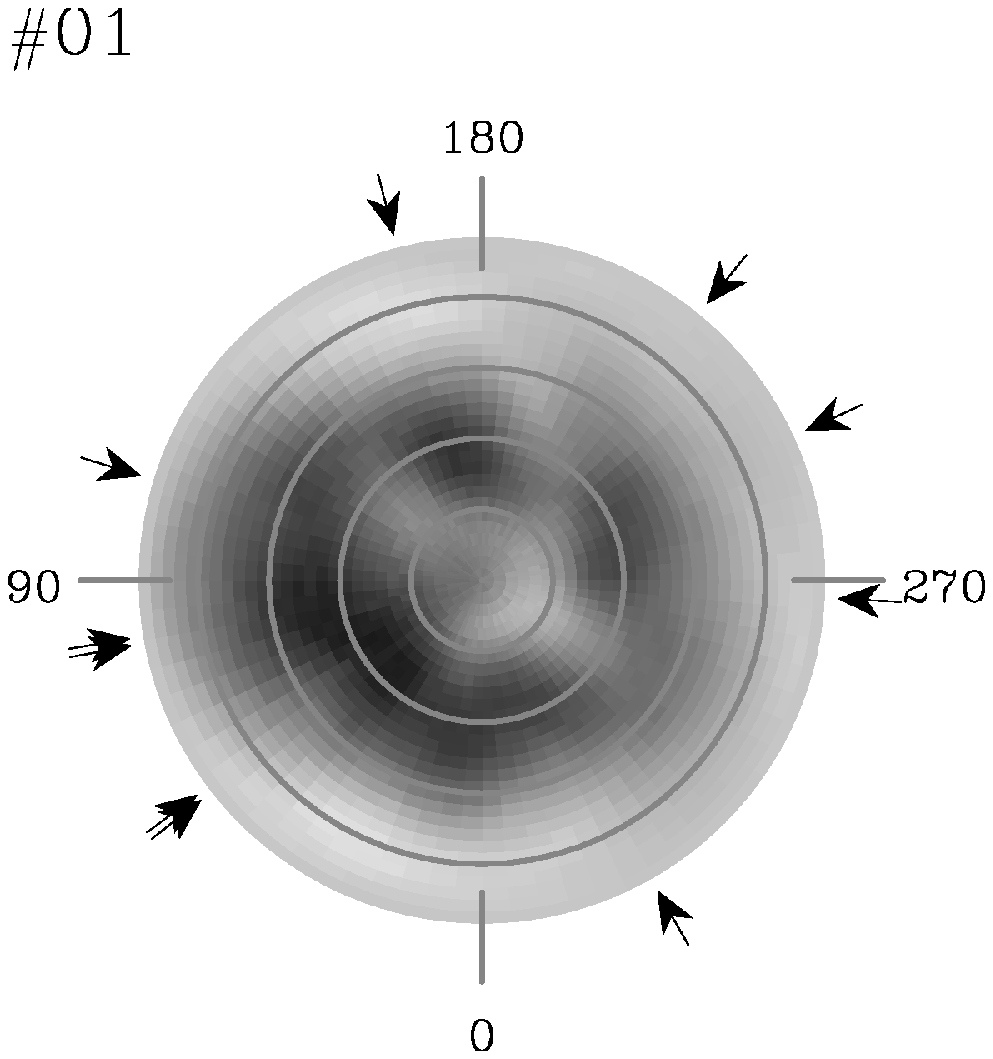}\hspace*{0. cm} \includegraphics[height=3.0cm]{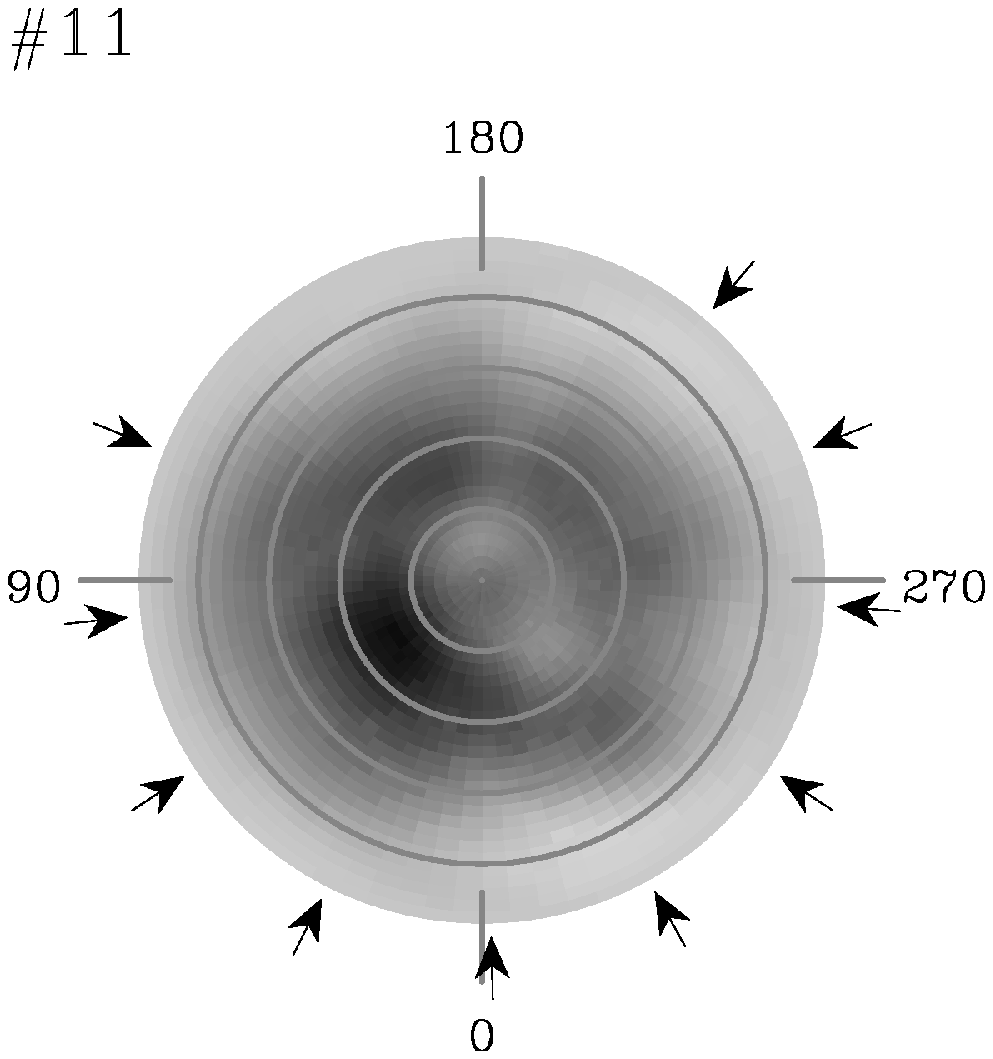} \hspace*{0. cm} \includegraphics[height=3.0cm]{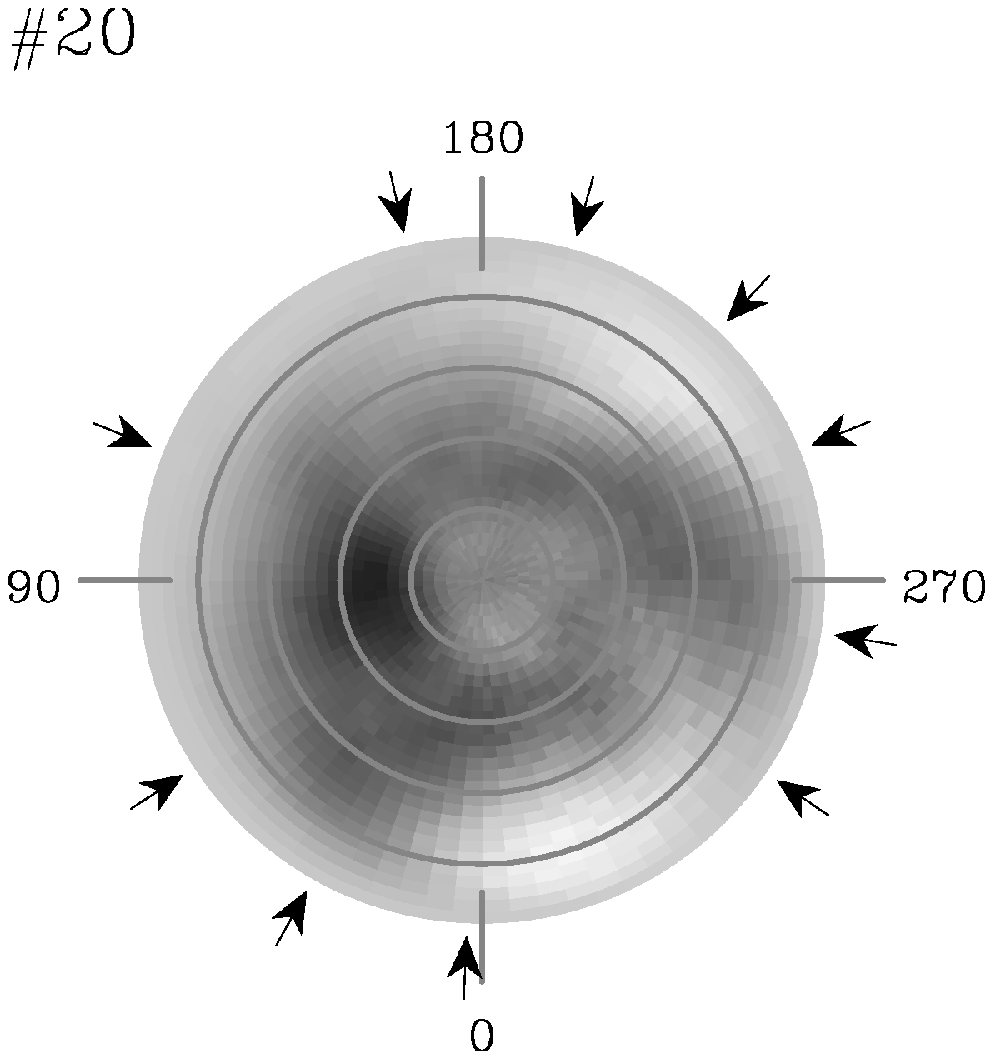} \hspace*{0. cm} \includegraphics[height=3.0cm]{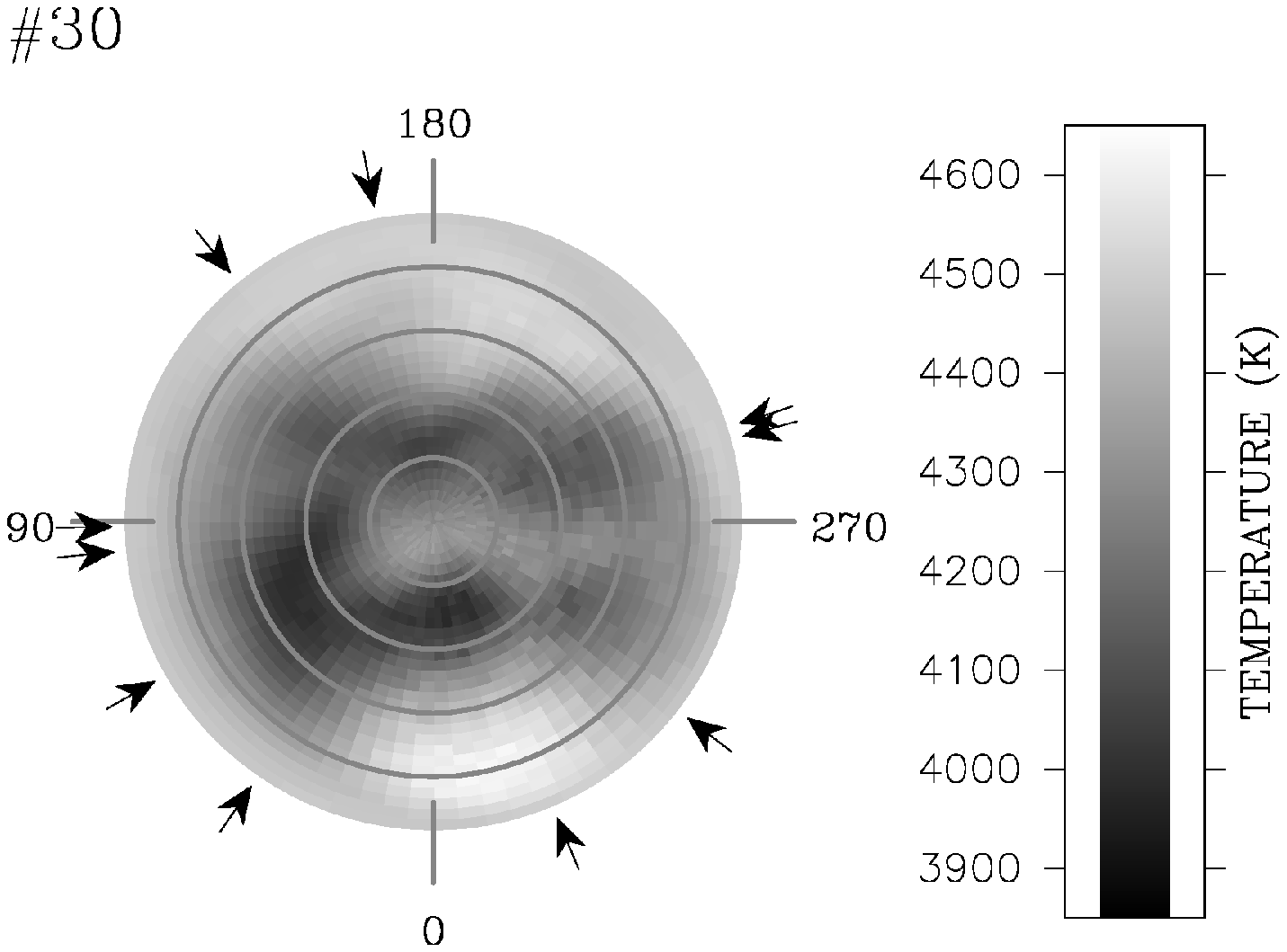} 
 \caption{Time evolution of the spotted surface of IL\,Hya from time-series Doppler imaging.}
   \label{fig1}
\end{center}
\end{figure}

IL\,Hya is a double-lined binary star (K0IV + G8V), a typical RS\,CVn-type system orbiting with a period of 12.905 days.
Our time-series spectroscopic dataset were obtained during a 70-night long observing run at NSO in 1996/97.
From that we reconstruct 30 time-series Doppler images for two favoured mapping lines (Fe\,{\sc i}-6430 and Ca\,{\sc i}-6439)
using our image reconstruction code {\sc TempMap} (\cite{Ricetal89}). Adopted astrophysical parameters
are listed in Table\,\ref{tab1}. As samples from the reconstructions, combined (Fe+Ca) maps
are shown in Fig.\,\ref{fig1}, indicating significant changes of the spotted surface over a few rotation cycles.

\begin{table}
  \begin{center}
  \caption{Astrophysical chart of IL\,Hya based on \cite[Weber \& Strassmeier (1998)]{WebStr98}}
  \label{tab1}
 {\scriptsize
  \begin{tabular}{ll}\hline 
Spectral type	&	K0IV (+G8V) \\
$\log g$	&	2.5$\pm$0.5 \\
$T_{\rm eff}$ [K]	&	4500$\pm$250 \\
$B-V$ [mag]	&	1.012$\pm$0.010 \\
$V-I$ [mag]	&	0.99$\pm$0.01 \\
Distance$^a$ [pc]	&	105.9$\pm$5.6 \\
$v\sin i$ [km\,s$^{-1}$]	&	26.5$\pm$1.0 \\
Inclination [deg]	&	55$\pm$5 \\
$P_{\rm orb}$ [days]	&	12.905$\pm$0.004 \\
Radius$^a$ [R$_\odot$]	&	8.1$\pm$0.9 \\
Microturbulence [km\,s$^{-1}$]	&	2.0 \\
Macroturbulence [km\,s$^{-1}$]	&	4.0 \\
Chemical abundances	&	0.9 dex below solar \\
Mass [M$_\odot$]	&	 $\approx$2.2 \\
\hline
$^a$ based on Hipparcos data &	\\
  \end{tabular}}
 \end{center}
\end{table}

\section{Surface differential rotation and meridional flow}

To measure surface DR we employ our method called `ACCORD' (acronym from Average Cross-CORrelation of consecutive Doppler images),
based on averaging cross-correlation function (ccf) maps of subsequent Doppler images. This way the surface differential rotation (hereafter DR) pattern in the ccf-maps could be
enhanced, while the unwanted effect of stochastic spot changes are supressed (see \cite[K\H{o}v\'ari et al. 2004,]{kovetal04} \cite[2007]{kovetal07} for details).
Applying ACCORD yields solar-type rotation law in the form of $\Omega(\beta) = \Omega_{\rm eq} - \Delta\Omega\sin^2\beta$ 
with an equatorial angular velocity $\Omega_{\rm eq}$ of 28.28$\pm$0.03 deg/day
and $\Delta\Omega = \Omega_{\rm eq} - \Omega_{\rm pole}$ of $-$1.43$\pm$0.15 deg/day,
corresponding with a surface shear $\alpha=\Delta\Omega / \Omega_{\rm eq}$ of $0.05\pm0.01$
(see the fitted average ccf-map in the left panel of Fig.\,\ref{fig1}). This shear is consistent with the
value of $\alpha=0.03\pm0.02$ derived by using a different method for a different dataset taken in 1988
(\cite[K\H{o}v\'ari \& Weber 2004]{KovWeb04}).
Regarding the reliability of the results read the other paper by K\H{o}v\'ari et al. in this proceedings.

\begin{figure}[h]
\begin{center}
 \includegraphics[width=4.5cm]{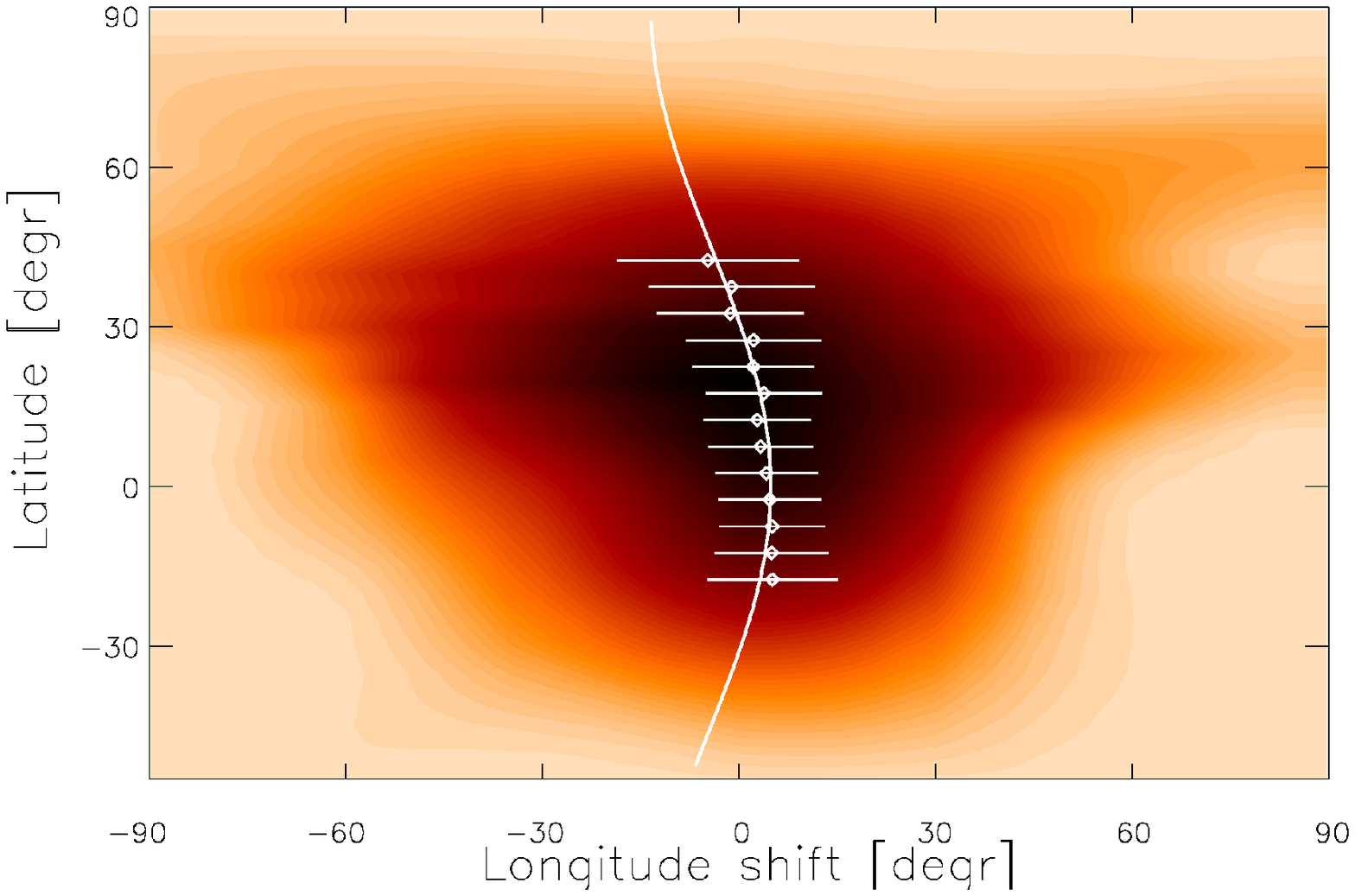}\hspace*{2.0 cm} \includegraphics[width=4.5cm]{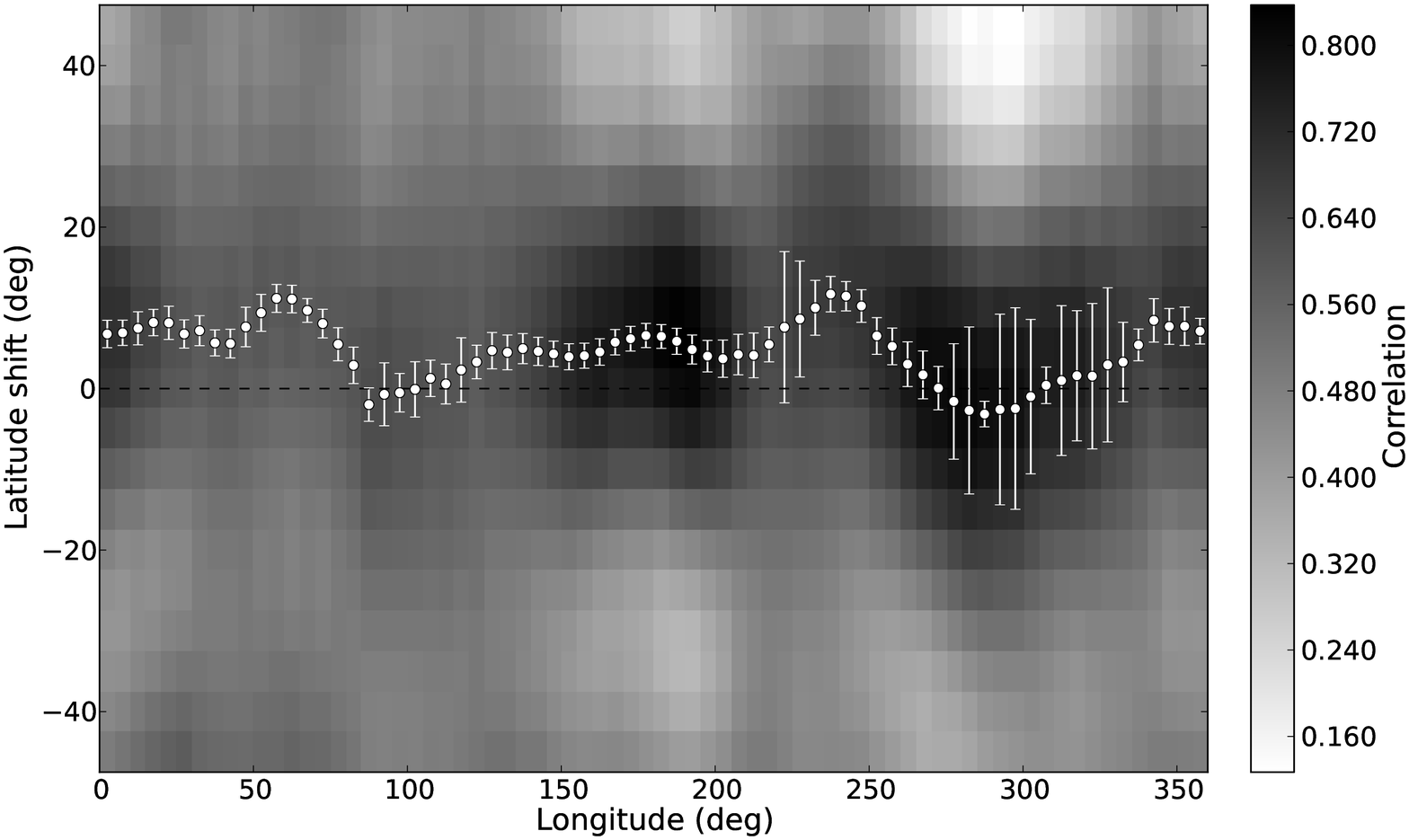}
 \caption{Averaged cross-correlations clearly reveal solar-type DR pattern (left) and common poleward drifting of spots (right).}
   \label{fig2}
\end{center}
\end{figure}

Latitudinal motion of spots can also be quantified by ACCORD. For this we use only the
hemisphere of the visible pole. For a detailed description of the method see \cite[K\H{o}v\'ari et al. (2007)]{kovetal07}. 
The resulting latitudinal correlation pattern (right panel in Fig.\,\ref{fig2})
can be converted into an average poleward surface velocity field of 0.5$\pm$0.1 km/s, that could be interpreted
as the surface pattern of a single-cell meridional circulation.

\begin{acknowledgments}
This work has been supported by the Hungarian Science Research Program OTKA K-81421,
the Lend\"ulet-2009 and Lend\"ulet-2012 Young Researchers' Programs of the Hungarian Academy
of Sciences and by the HUMAN MB08C 81013 grant of the MAG Zrt.
\end{acknowledgments}

\end{document}